\documentclass[aps,showpacs,twocolumn]{revtex4}
\usepackage{amssymb}
\usepackage{amsmath}

\begin{document}

\title{Dark matter as a geometric effect in $f(R)$ gravity}

\author{Christian~G.~B\"ohmer}
\email{c.boehmer@ucl.ac.uk} \affiliation{Department of
Mathematics, University College London,
             Gower Street, London, WC1E 6BT, UK}
\affiliation{Institute of Cosmology \& Gravitation,
             University of Portsmouth, Portsmouth PO1 2EG, UK}

\author{Tiberiu~Harko}
\email{harko@hkucc.hku.hk} \affiliation{Department of Physics and
Center for Theoretical
             and Computational Physics, The University of Hong Kong,
             Pok Fu Lam Road, Hong Kong}

\author{Francisco S.~N.~Lobo}
\email{francisco.lobo@port.ac.uk} \affiliation{Institute of
Cosmology \& Gravitation,
             University of Portsmouth, Portsmouth PO1 2EG, UK}
\affiliation{Centro de Astronomia e Astrof\'{\i}sica da
             Universidade de Lisboa, Campo Grande, Ed. C8 1749-016 Lisboa,
             Portugal}

\date{\today}

\begin{abstract}
We consider the behavior of the tangential velocity of test particles moving
in stable circular orbits in $f(R)$ modified theories of gravity. A large
number of observations at the galactic scale have shown that the rotational
velocities of massive test particles (hydrogen clouds) tend towards constant
values at large distances from the galactic center. We analyze the vacuum
gravitational field equations in $f(R)$ models in the constant velocity
region, and the general form of the metric tensor is derived in a closed
form. The resulting modification of the Einstein-Hilbert Lagrangian is of
the form $R^{1+n}$, with the parameter $n$ expressed in terms of the
tangential velocity. Therefore we find that to explain the motion of test
particles around galaxies requires only very mild deviations from classical
general relativity, and that modified gravity can explain the galactic
dynamics without the need of introducing dark matter.
\end{abstract}


\maketitle

\section{Introduction}

Modern astrophysical and cosmological models are faced with two severe
theoretical difficulties, that can be summarized as the dark energy and the
dark matter problems. Despite the fact that several suggestions have
recently been proposed to overcome these issues, a satisfactory answer is
yet to be obtained. However, in the context of dark matter, two
observations, namely, the behavior of the galactic rotation curves and the
mass discrepancy in galactic clusters, suggest the existence of a (non or
weakly interacting) form of dark matter at galactic and extra-galactic
scales.

The galactic rotation curves of spiral galaxies~\cite{Bi87} are probably the
most striking evidences for the possible failure of Newtonian gravity and of
the general theory of relativity on galactic and intergalactic scales. In
these galaxies, neutral hydrogen clouds are observed at large distances from
the center, much beyond the extent of the luminous matter. As these clouds
are moving in circular orbits with nearly constant tangential velocity $v_{%
\mathrm{tg}}$, such orbits are maintained by the balance between the
centrifugal acceleration $v_{\mathrm{tg}}^{2}/r$ and the gravitational
attraction $GM(r)/r^{2}$ of the total mass $M(r)$ contained within the
radius $r$. This yields an expression for the galactic mass profile of the
form $M(r)=rv_{\mathrm{tg}}^{2}/G$, with the mass increasing linearly with $r
$, even at large distances, where very little luminous matter has been
detected~\cite{Bi87}. This peculiar behavior of the rotation curves is
usually explained by postulating the existence of dark matter, assumed to be
a cold and pressureless medium, distributed in a spherical halo around the
galaxies.

There are many possible candidates for dark matter, the most popular ones
being the weakly interacting massive particles (WIMP). Their interaction
cross sections with normal baryonic matter, although extremely small, are
expected to be non-zero, and therefore it is believed to detect them
directly~\cite{OvWe04}. However, no direct (non-gravitational) evidence for
the existence of dark matter has been reported so far. It is important to
emphasize that dark matter consisting of WIMP's may exist in the form of an
Einstein cluster~\cite{Ein}, or could possibly undergo a phase transition to
form a Bose-Einstein condensate~\cite{Bo}. One cannot also \textit{a priori}
exclude the possibility that Einstein's (and Newton's) theory of gravity
breaks down at galactic scales. In this context, several theoretical models,
based on a modification of Newton's law or of general relativity, have been
proposed so far to explain the behavior of the galactic rotation curves~\cite
{expl1,expl2,expl5}.

A promising avenue that has been extensively investigated recently are the $%
f(R)$ modified theories of gravity, where the standard
Einstein-Hilbert action is replaced by an arbitrary function of
the Ricci scalar $R$~\cite {first}. In this work we shall use the
metric formalism, which consists in varying the action with
respect to $g_{\mu\nu}$, although other alternative approaches
have been considered in the literature, namely, the Palatini
formalism~\cite{Palatini,Sotiriou:2006qn}, where the metric and
the connections are treated as separate variables; and the
metric-affine formalism, where the matter part of the action now
depends and is varied with respect to the
connection~\cite{Sotiriou:2006qn}. It has been suggested that
these modified gravity models account for the late time
acceleration of the universe~\cite{Carroll:2003wy}, thus
challenging the need for dark energy. However, the viability of
the $f(R)$ models proposed in the literature has been extensively
analyzed~\cite {viablemodels,Hu:2007nk,energycond}. In this
context, severe weak field constraints in the solar system range
seem to rule out most of the models proposed so
far~\cite{solartests,Zakharov:2006uq,Chiba,Fa08}, although viable
models do exist~\cite
{Hu:2007nk,solartests2,Sawicki:2007tf,Amendola:2007nt}.

In order to be a viable theory, in addition to satisfying the solar system
constraints, the proposed models should simultaneously account for the four
distinct cosmological phases, namely, inflation, the radiation-dominated and
matter-dominated epochs, and the late-time accelerated expansion~\cite
{Nojiri:2006be}, and be consistent with cosmological structure formation
observations~\cite{structureform}. The issue of stability~\cite
{Faraoni:2006sy} also plays an important role in the viability of
cosmological solutions~\cite{Nojiri:2006ri,Amendola:2007nt,Sawicki:2007tf}.
It is interesting to note that, recently, viable cosmological $f(R)$ models
were analyzed, and it was found that the latter models satisfying
cosmological and local gravity constraints are practically indistinguishable
from the $\Lambda$CDM model, at least at the background level~\cite
{Amendola:2007nt}.

The possibility that the galactic dynamics of massive test
particles may be understood without the need for dark matter was
also considered in the framework of $f(R)$ modified theories of
gravity \cite{Cap2,Borowiec:2006qr,Mar1}. In the context of
galactic dynamics, a version of $f(R)$ gravity models admitting a
modified Schwarzschild-de Sitter metric was analyzed in~\cite
{Sob}. In the weak field limit one obtains a small logarithmic
correction to the Newtonian potential, and a test star moving in
such a spacetime acquires a constant asymptotic speed at large
distances. It is interesting to note that the model has similar
properties with MOND~\cite{expl1}. A model based on a generalized
action with $f(R) = R + R(R/R_0 + 2/\alpha)^{-1}\ln(R/R_c)$, where
$\alpha $, $R_0$ and $R_c$ are constants, was proposed
in~\cite{SaRa07}. In particular, this model can describe the
Pioneer anomaly and the flat rotation curves of the spiral
galaxies. In a cosmological context, the vacuum solution also
results in a late time acceleration for the universe. The
generalization of the virial theorem in $f(R)$ modified gravity,
using the collisionless Boltzmann equation, was considered in
\cite{BoHaLo08}.

Within the framework of $f(R)$ gravity, a model exhibiting an explicit
coupling of an arbitrary function of $R$ with the matter Lagrangian density
was proposed recently~\cite{Bertolami:2007gv}. Due to this coupling a
connection between the problem of the rotation curve of galaxies, via a
solution somewhat similar to the one put forward in the context of MOND, and
the Pioneer anomaly is established.

It is the purpose of the present paper to consider, from an exact analytic
point of view, the problem of the galactic rotation curves in the framework
of $f(R)$ modified theories of gravity. In order to find an exact analytic
description of the galactic dynamics of test particles in $f(R)$ gravity
models, we start from the general relativistic expression of the tangential
velocity $v_{\mathrm{tg}}$ of massive test particles in static and
spherically symmetric spacetimes, moving in stable circular orbits around
the galactic center. The rotational velocity is determined by the $g_{tt}$
component of the metric tensor and of the radial distance only.

We limit our analysis to the most important region for the galactic dynamics
of test particles, namely, the region of constant rotational velocities. The
constancy of the tangential velocity completely determines the form of $%
g_{tt}$, and consequently, the exact analytical form of the latter metric
tensor component is completely determined from dynamical considerations.

As a next step in our analysis of the geometry in the constant velocity
region we consider the spherically symmetric vacuum solutions of the
gravitational field equations in $f(R)$ modified theories of gravity. By
introducing several coordinate and functional transformations, the field
equations can be reduced to an autonomous system of differential equations.
By using the general form of $g_{tt}$ we obtain a second order differential
equation fixing the functional form of $g_{rr}$ in the constant velocity
region. This equation has as an exact solution $g_{rr}=\mathrm{constant}$,
which gives $g_{rr}$ as a function of the observed tangential velocity only.
The expressions of the Ricci scalar and of the function $f(R)$ are also
obtained.

As a general conclusion of our study we find that to explain the flat
galactic rotation curves, only small deviations from standard general
relativity are needed, so that $f(R)\propto R^{1+v_{\mathrm{tg}}^{2}}$, a
somewhat natural result in the context of modified gravity theories.

As a possible observational test of our results we suggest the study of the
lensing of light by galaxies in the constant velocity region. The study of
lensing may, in principle, discriminate between the present model and other
dark matter models. On the other hand, the deflection angle in our model is
of the same order of magnitude as the deflection angle in the standard
isothermal sphere dark matter model, which is well tested observationally.
This shows that the results obtained in this paper are consistent both
theoretically and observationally.

The present paper is organized as follows. The tangential velocity of a test
particle in modified theories of gravity is derived in Section \ref{Sec:II}.
The vacuum field equations in the $f(R)$ models are written down in Section
\ref{Sec:III}. Two specific solutions of the field equations in the constant
tangential velocity region are presented in Section \ref{geom}, where some
general properties of the basic equation describing the behavior of $g_{rr}$
are also discussed. We discuss and conclude our results in Section \ref
{Sec:IV}.

\section{The motion of massive test particles in stable circular orbits}

\label{Sec:II}

In order to obtain results which are relevant to the galactic dynamics, in
the following, we restrict our study to the static and spherically symmetric
metric given by
\begin{align}
ds^{2}=-e^{\nu (r)}dt^{2}+e^{\lambda (r)}dr^{2}+r^{2}d\Omega ^{2},
\label{metr1}
\end{align}
where $d\Omega ^{2}=d\theta ^{2}+\sin ^{2}\theta d\phi ^{2}$. In the present
paper, we use a system of units so that $G=c=1$. The metric tensor
coefficient $e^{\lambda (r)}$ is constrained by the condition $e^{\lambda
(r)}\geq 1$, $\forall r\in \lbrack 0,\infty )$, while we assume that $e^{\nu
(r)}$ satisfies the condition $e^{\nu (r)}\geq 0$, $\forall r\in \lbrack
0,\infty ) $.

The Lagrangian $\mathcal{L}$ for a massive test particle reads
\begin{align}
\mathcal{L}=\frac{1}{2}\left( -e^{\nu }\dot{t}^{2}+e^{\lambda }\dot{r}%
^{2}+r^{2}\dot{\Omega}^{2}\right) ,  \label{lag}
\end{align}
where the overdot denotes differentiation with respect to the affine
parameter $s$. Since the metric tensor coefficients do not explicitly depend
on $t$ and $\Omega $, the Lagrangian~(\ref{lag}) yields the following
conserved quantities (generalized momenta)~\cite{Matos}:
\begin{align}
-e^{\nu (r)}\dot{t}=E,\qquad r^{2}\dot{\Omega}=L,  \label{cons}
\end{align}
where $E$ is related to the total energy of the particle and $L$ to the
total angular momentum. With the use of the conserved quantities, we obtain
from Eq.~(\ref{lag}) the geodesic equation for massive particles in the form
\begin{align}
e^{\nu +\lambda }\dot{r}^{2}+e^{\nu }\left( 1+\frac{L^{2}}{r^{2}}\right)
=E^{2}\,.  \label{geod1}
\end{align}

For the case of the motion of particles in circular and stable orbits the
effective potential must satisfy the following conditions: a) $\dot{r}=0$,
representing circular motion; b) $\partial V_{eff}/\partial r$ $=0$,
providing extreme motion; c) $\partial ^{2}V_{eff}/\partial r$ $^{2}|
_{extr}>0$, translating a stable orbit~\cite{Matos}. Conditions a) and b)
immediately provide the conserved quantities as
\begin{align}
E^{2}=e^{\nu }\left( 1+\frac{L^{2}}{r^{2}}\right) ,  \label{cons1}
\end{align}
and
\begin{align}
\frac{L^{2}}{r^{2}}=\frac{r\nu ^{\prime }}{2}e^{-\nu }E^{2},  \label{cons2}
\end{align}
respectively. Equivalently, these two equations can be rewritten as
\begin{align}
E^{2}=\frac{e^{\nu }}{1-r\nu ^{\prime }/2}\,,\qquad L^{2}=\frac{r^{3}\nu
^{\prime }/2}{1-r\nu ^{\prime }/2}\,.
\end{align}

We define the tangential velocity $v_{\mathrm{tg}}$ of a test particle, as
measured in terms of the proper time~\cite{LaLi}, that is, by an observer
located at the given point, as
\begin{align}
v_{\mathrm{tg}}^{2}=e^{-\nu }r^{2}\left( \frac{d\Omega }{dt}\right)
^{2}=e^{-\nu }r^{2}\dot{\Omega}^{2}/\dot{t}^{2}=e^{\nu }\frac{L^{2}}{%
r^{2}E^{2}}\,.  \label{vtgbr}
\end{align}
By using the constants of motion, we obtain the expression of the tangential
velocity of a test particle in a stable circular orbit~\cite{Matos}, given
by
\begin{align}
v_{\mathrm{tg}}^{2}=\frac{r\nu ^{\prime }}{2}.  \label{vtg}
\end{align}

This simple expression which relates one of the metric components to the
tangential velocity has three important properties. First of all, it is an
exact general relativistic expression valid for static and spherically
symmetric spacetimes. Secondly, it is interesting to note that the
tangential velocity is sensitive to only one of the two metric functions,
i.e., it is independent of the form of $g_{rr}$. Lastly, since the motion of
test particles is defined via the geodesic equations, this relation is
independent of $f(R)$. Even if we allow the modified theory of gravity to
contain arbitrary contractions of the Ricci and Riemann tensors, the above
equations would still hold exactly.

In regions with constant tangential velocity the metric function $\nu $ is
fixed by the condition $v_{\mathrm{tg}}\approx $ constant, and by
integrating Eq.~(\ref{vtg}), we find
\begin{align}
\nu =2v_{\mathrm{tg}}^{2}\ln \left( \frac{r}{r_{0}}\right),  \label{nu}
\end{align}
where $r_{0}$ is a constant of integration. Therefore the most general
static and spherically symmetric metric in the constant tangential velocity
regions can be written as
\begin{align}
ds^{2}=-\left( \frac{r}{r_{0}}\right) ^{2v_{\mathrm{tg}}^{2}}dt^{2}+e^{%
\lambda (r)}dr^{2}+r^{2}d\Omega ^{2}.  \label{nu1}
\end{align}

Note, however, that the galactic rotation curves generally show a more
complicated dynamics~\cite{Bi87}. Therefore, to obtain a more accurate and
realistic description of the particle's motion, which can be extended beyond
the constant velocity region, more general expressions of the metric
coefficients are needed. However, in the present paper we restrict our
analysis to the constant velocity region, which provides interesting results.

In order to test the consistency of our results, we consider the Newtonian
limit of the model. The assumption of small velocities of the particles
requires that the gravitational field be weak. In the Newtonian limit the $%
g_{tt}$ component of the metric tensor is given by $e^{\nu}\approx
1+2\Phi_{N}$, where $\Phi _{N}$ is the Newtonian gravitational potential
satisfying the Poisson equation $\Delta \Phi_{N}=4\pi\rho$~\cite{LaLi}. In
the constant velocity region the mass $M(r)$ of the dark matter and the
energy density $\rho$ vary with the distance as $M(r)=v_{\mathrm{tg}}^{2}r$
and $\rho =v_{\mathrm{tg}}^{2}/4\pi r^{2}$, respectively. Therefore the
Poisson equation is given by
\begin{align}
\frac{1}{r^{2}}\frac{d}{dr}\left( r^{2}\frac{d\Phi _{N}}{dr}\right) =\frac{%
v_{\mathrm{tg}}^{2}}{r^{2}},
\end{align}
and has the general solution
\begin{align}
\Phi _{N}(r)=v_{\mathrm{tg}}^{2}\ln \frac{r}{r_{0}}-\frac{C_{N}}{r},
\end{align}
where $C_{N}$ and $r_{0}$ are arbitrary constants of integration. In the
limit of large $r$, corresponding to the constant velocity regions around
galaxies, the Newtonian potential is given by
\begin{align}\label{logpot}
\Phi _{N}(r)\approx v_{\mathrm{tg}}^{2}\ln \frac{r}{r_{0}},
\end{align}
reflecting a logarithmic dependence on the radial distance $r$. On the other
hand, the $g_{tt}$ component of the metric tensor can be represented in the
constant velocity ``dark matter'' region as
\begin{align}
e^{\nu } &\approx \left( \frac{r}{r_{0}}\right) ^{2v_{\mathrm{tg}}^{2}}=\exp %
\left[ \ln \left( \frac{r}{r_{0}}\right) ^{2v_{\mathrm{tg}}^{2}}\right]
\notag \\
& \approx 1+2v_{\mathrm{tg}}^{2}\ln \left( \frac{r}{r_{0}}\right)
=1+2\Phi_{N}(r).
\end{align}

Therefore the model has a well-defined Newtonian limit, and the metric given
by Eq.~(\ref{nu1}) can indeed be used to describe the geometry of the
spacetime in the dark matter dominated regions.

\section{Vacuum field equations in $f(R)$ gravity}

\label{Sec:III}

The action for the modified theories of gravity considered in this work
takes the following form
\begin{align}
S=\int f(R)\sqrt{-g}\;d^{4}x,
\end{align}
where $f(R)$ is an arbitrary analytical function of the Ricci scalar $R$.
Note that we are only interested in the vacuum case, and therefore we have
not added a matter Lagrangian to the action.

Varying the action with respect to the metric $g_{\mu \nu }$ yields the
following field equations
\begin{align}
F(R)R_{\mu \nu }-\frac{1}{2}f(R)g_{\mu \nu }-\left(\nabla _{\mu }\nabla
_{\nu }-g_{\mu \nu }\square \right)F(R)=0,  \label{field}
\end{align}
where we have denoted $F(R)=d f(R)/d R$. Note that the covariant derivative
of these field equations vanishes for all $f(R)$ by means of the generalized
Bianchi identities~\cite{Bertolami:2007gv,Koivisto}.

For a static and spherically symmetric metric of the form given by Eq.~(\ref
{metr1}), the field equations of the $f(R)$ gravity in vacuum can be
expressed as~\cite{Sob,Multamaki:2006zb}
\begin{align}
F^{\prime \prime }-\frac{1}{2}\left( \nu ^{\prime }+\lambda ^{\prime
}\right) F^{\prime }-\frac{\left( \nu ^{\prime }+\lambda ^{\prime }\right) }{%
r}F=0,  \label{f1}
\end{align}
\begin{multline}
\nu ^{\prime \prime }+\nu ^{\prime 2}-\frac{1}{2}\left( \nu ^{\prime
}+\lambda ^{\prime }\right) \left( \nu ^{\prime }+\frac{2}{r}\right) -\frac{2%
}{r^{2}}\left( 1-e^{\lambda }\right) \\
=-2\frac{F^{\prime \prime }}{F}+\left( \lambda ^{\prime }+\frac{2}{r}\right)
\frac{F^{\prime }}{F},  \label{f2}
\end{multline}
\begin{align}
f=Fe^{-\lambda }\left[ \nu ^{\prime \prime }-\frac{1}{2}\left( \nu ^{\prime
}+\lambda ^{\prime }\right) \nu ^{\prime }-\frac{2}{r}\lambda ^{\prime
}+\left( \nu ^{\prime }+\frac{4}{r}\right) \frac{F^{\prime }}{F}\right] ,
\label{f3}
\end{align}
\begin{align}
R=2\frac{f}{F}-3e^{-\lambda }\left\{ \frac{F^{\prime \prime }}{F}+\left[
\frac{1}{2}\left( \nu ^{\prime }-\lambda ^{\prime }\right) +\frac{2}{r}%
\right] \frac{F^{\prime }}{F}\right\} .  \label{f4}
\end{align}

It is useful to introduce a new variable $\eta $ by means of the following
transformation
\begin{align}
\eta =\ln r.
\end{align}

Therefore, the field equations Eqs.~(\ref{f1})--(\ref{f4}) take the form
\begin{align}
\frac{d^{2}F}{d\eta ^{2}}-\left[ 1+\frac{1}{2}\left( \frac{d\nu }{d\eta }+%
\frac{d\lambda }{d\eta }\right) \right] \frac{dF}{d\eta }-\left( \frac{d\nu
}{d\eta }+\frac{d\lambda }{d\eta }\right) F=0,  \label{g1}
\end{align}
\begin{multline}
\frac{d^{2}\nu }{d\eta ^{2}}-\frac{d\nu }{d\eta }+\left( \frac{d\nu }{d\eta }%
\right) ^{2}-\frac{1}{2}\left( \frac{d\nu }{d\eta }+\frac{d\lambda }{d\eta }%
\right) \left( \frac{d\nu }{d\eta }+2\right) \\
+2\left( 1-e^{\lambda}\right) = -2\frac{1}{F}\frac{d^{2}F}{d\eta ^{2}}%
+\left( \frac{d\lambda }{d\eta }+4\right) \frac{1}{F}\frac{dF}{d\eta },
\label{g2}
\end{multline}
\begin{multline}
f=Fe^{-\lambda -2\eta }\biggl[ \frac{d^{2}\nu }{d\eta ^{2}}-\frac{d\nu }{%
d\eta }-\frac{1}{2}\left( \frac{d\nu }{d\eta }+\frac{d\lambda }{d\eta }%
\right) \frac{d\nu }{d\eta } \\
-2\frac{d\lambda }{d\eta }+\left( \frac{d\nu }{d\eta }+4\right) \frac{1}{F}%
\frac{dF}{d\eta }\biggr] ,  \label{g3}
\end{multline}
\begin{multline}
R=2\frac{f}{F}-3e^{-\lambda -2\eta }\biggl\{ \frac{1}{F}\frac{d^{2}F}{d\eta
^{2}}-\frac{1}{F}\frac{dF}{d\eta } \\
+\left[ \frac{1}{2}\left( \frac{d\nu }{d\eta }-\frac{d\lambda }{d\eta }%
\right) +2\right] \frac{1}{F}\frac{dF}{d\eta }\biggr\} .  \label{g4}
\end{multline}

As a result of introducing the new variable the basic field equations Eqs.~(%
\ref{g1}) and~(\ref{g2}) are independent of the radial coordinate $\eta $.
It is also very useful to introduce a formal representation of the function $%
F$ as
\begin{align}
F\left( \eta \right) =F_{0}\exp \left[ \int u(\eta )d\eta \right] ,
\label{sol1}
\end{align}
where $u$ is a new function of $\eta $, and $F_{0}$ is an arbitrary
constant, so that $(1/F)dF/d\eta =u$, $(1/F)d^{2}F/d\eta ^{2}=du/d\eta +u^{2}
$. Hence Eq.~(\ref{g1}) can be written as
\begin{align}
\frac{du}{d\eta }+u^{2}-\left[ 1+\frac{1}{2}\left( \frac{d\nu }{d\eta }+%
\frac{d\lambda }{d\eta }\right) \right] u-\left( \frac{d\nu }{d\eta }+\frac{%
d\lambda }{d\eta }\right) =0.  \label{h1}
\end{align}

This equation is a Riccati type first order differential equation. To find
its general solution the knowledge of a particular solution is required. By
using the function $u$, Eq.~(\ref{g2}) can be written as
\begin{multline}
\frac{d^{2}\nu }{d\eta ^{2}}-\frac{d\nu }{d\eta }+\left( \frac{d\nu }{d\eta }%
\right) ^{2}-\frac{1}{2}\left( \frac{d\nu }{d\eta }+\frac{d\lambda }{d\eta }%
\right) \left( \frac{d\nu }{d\eta }+2\right) \\
+2\left(1-e^{\lambda }\right) = -2\frac{du}{d\eta }-2u^{2}+\left( \frac{%
d\lambda }{d\eta }+4\right) u.  \label{h2}
\end{multline}

Substituting the term $du/d\eta +u^{2}$ in Eq.~(\ref{h2}), with the use of
Eq.~(\ref{h1}), we obtain
\begin{multline}
2\left( 1-\frac{1}{2}\frac{d\nu }{d\eta }\right) u=2\left( 1-e^{\lambda
}\right) +\frac{d^{2}\nu }{d\eta ^{2}}-\frac{d\nu }{d\eta }+\left( \frac{%
d\nu }{d\eta }\right) ^{2} \\
+\left( \frac{d\nu }{d\eta }+\frac{d\lambda }{d\eta }\right) \left( 1-\frac{1%
}{2}\frac{d\nu }{d\eta }\right).  \label{sol2}
\end{multline}

The general solution of Eqs.~(\ref{h1}) and~(\ref{sol2}) provides the
general static and vacuum solution of the modified field equations for the
case of the $f(R)$ gravity models. Once $\nu (\eta )$ and $\lambda (\eta )$
are specified, one can immediately obtain $u$ and consequently (by
integration) $F$, as well as all the other relevant physical quantities. If
the function $F$ and the metric tensor coefficients are known, $f$ can be
obtained as a function of $R$ from Eqs.~(\ref{g3}) and~(\ref{g4}) in a
parametric form, as $f=f(\eta) $, $R=R(\eta) $.

\section{Dark matter as a geometric effect in $f(R)$ models}

\label{geom}

In order to obtain a geometric interpretation of dark matter in $f(R) $
gravity, we start with the metric coefficients $\nu $ and $\lambda $ in the
dark matter dominated region at the galactic scale, given by Eq.~(\ref{nu1}%
), which, using the variable $\eta $, can be represented in the following
form
\begin{align}
\nu =2m\left( \ln \eta -\ln \eta _{0}\right),\qquad \lambda =\lambda \left(
\eta\right) ,
\end{align}
where $\eta _{0}=\ln r_{0}=\mathrm{constant}$, and for notational simplicity
we have denoted
\begin{align}
m=v_{\mathrm{tg}}^{2}=\mathrm{constant}.
\end{align}

For this form of the metric Eqs.~(\ref{h1}) and~(\ref{sol2}) become
\begin{align}
\frac{du}{d\eta }+u^{2}-\left( 1+m+\frac{1}{2}\frac{d\lambda }{d\eta }%
\right) u-2m-\frac{d\lambda }{d\eta }=0,  \label{eqf1}
\end{align}
and
\begin{align}
2\left( 1-m\right) u=2\left( 1-e^{\lambda }\right) +2m^{2}+\left( 1-m\right)
\frac{d\lambda }{d\eta },  \label{eqf2}
\end{align}
respectively. By taking the derivative with respect to $\eta $ of Eq.~(\ref
{eqf2}) and substituting the resulting $du/d\eta $ and $u$ in Eq.~(\ref{eqf1}%
) provides the following second order differential equation
\begin{multline}
\frac{1}{2}\frac{d^{2}\lambda }{d\eta ^{2}}-\frac{1}{1-m}\left( \frac{3}{2}%
e^{\lambda }-m^{2}+1-m\right) \frac{d\lambda }{d\eta} \\
+\frac{1}{\left( 1-m\right) ^{2}}\left( 1-e^{\lambda }\right)^{2}+ \frac{%
3m^{2}-1}{\left( 1-m\right) ^{2}}\left( 1-e^{\lambda }\right) \\
+\frac{m^{2}\left( 2m^{2}-1\right) }{\left( 1-m\right)^{2}}-2m=0\,,
\label{eqlambda}
\end{multline}
which must be satisfied by the metric coefficient $\lambda(\eta)$ in the
``dark matter'' dominated regions in $f(R)$ gravity models.

From astrophysical observations it is known that the tangential velocity of
test particles in circular stable orbits around the galactic center is of
the order of $v_{\mathrm{tg}}\approx 200-300 \;\mathrm{km}/\mathrm{s}$~\cite
{Bi87}. Hence, we have $m=v_{\mathrm{tg}}^2\approx 10^{-6}$, and all the
terms containing $m^{2}$ and $m^4$ in Eq.~(\ref{eqlambda}) can be neglected
within a very good approximation. Considering, thus, a first order
approximation in $m$, Eq.~(\ref{eqlambda}) describing the metric coefficient
$\exp(\lambda)$ reduces to
\begin{multline}
\frac{1}{2}\frac{d^{2}\lambda }{d\eta ^{2}}-\frac{1}{1-m}\left( \frac{3}{2}%
e^{\lambda }+1-m\right) \frac{d\lambda }{d\eta }+\frac{1}{\left(
1-m\right)^{2}} \\
\times\left( 1-e^{\lambda }\right)^{2} -\frac{1}{\left( 1-m\right) ^{2}}%
\left( 1-e^{\lambda }\right) -2m=0.  \label{eqlambda1}
\end{multline}

In the following analysis we will consider two classes of solutions of Eq.~(%
\ref{eqlambda1}).

\subsection{The case of the constant $\protect\lambda $}

Equation~(\ref{eqlambda1}) admits an exact solution of the form $\lambda =%
\mathrm{constant}$. By denoting $1-e^{\lambda }=\delta $, it follows that
considering a first order approximation in $m$, then $\delta $ satisfies the
following second order algebraic equation
\begin{align}
\delta ^{2}-\delta -2m=0,
\end{align}
which has only one physical solution, namely, $\delta =-2m$ (the other
solution contradicts the condition $e^{\lambda }\geq 1$, which essentially
represents the positivity of the mass). Therefore, in the first order
approximation the metric coefficient $e^{\lambda }$ in the ``dark matter''
region becomes
\begin{align}
e^{\lambda }\approx 1+2v_{\mathrm{tg}}^{2},\qquad e^{-\lambda }\approx 1-2v_{%
\mathrm{tg}}^{2}.  \label{rez1}
\end{align}

Equation~(\ref{rez1}) has a straightforward physical interpretation. Since
in the Newtonian approximation $v_{\mathrm{tg}}^{2}=GM(r)/r$, where $M(r)$
is the total mass of the galaxy, we obtain the metric in a form which is
very similar to the Schwarzschild solution of general relativity, i.e., $%
e^{-\lambda }\approx 1-2GM(r)/r$. On the other hand, since $v_{\mathrm{tg}%
}^{2}=\mathrm{constant}$, the mass within the radius $r$ must increase so
that $M(r)\sim r$. This ``mass'', which is linearly increasing with the
distance, is usually interpreted as due to the presence of the dark matter.

In the present approach, the effect of the appearance of a dark ``mass'' is
of a purely geometric origin, resulting from the modification of the basic
equations of the gravitational field. Nevertheless, one can formally define
a mass $M(r)$ for the dark matter, despite the fact that the origin of this
``mass'' cannot be related to physical particles. Thus, the rotational
galactic curves can be naturally explained in $f(R)$ gravity models without
introducing any additional hypothesis. The galaxy is embedded in a modified
spherically symmetric geometry, generated by the non-zero contributions of
the modified gravitational action. The extra-terms act as a ``matter''
distribution outside the galaxy.

Once the metric coefficients are known, Eq.~(\ref{eqf1}) provides the
function $u(\eta)$ to first order in $m$ as
\begin{align}
u\left( \eta \right) =\frac{1+m}{2}+\frac{\sqrt{1+10m}}{2}\tanh \left[ \frac{%
\sqrt{1+10m}\left( \eta -\eta _{1}\right) }{2}\right] ,
\end{align}
where $\eta _{1}$ is an arbitrary integration constant. The function $F(\eta
)$ can be found as
\begin{align}
F\left( \eta \right) =F_{0}\exp \left( \frac{1+m}{2}\eta \right) \cosh \left[
\frac{\sqrt{1+10m}\left( \eta -\eta _{1}\right) }{2}\right] .
\end{align}

For large $r$, by taking one arbitrary integration constant equal to zero,
without a significant loss of generality, so that we keep only the
decreasing term in the expression of $F$, we obtain
\begin{align}
F(r)=C_{1}r^{-2m},
\end{align}
where $C_{1}$ is an arbitrary constant. Finally, from Eqs.~(\ref{f3}) and~(%
\ref{f4}) we find
\begin{align}  \label{n}
f(R)=f_{0}R^{1+m},
\end{align}
where $f_{0}$ is a constant.

\subsection{First order corrections and general properties of $\protect%
\lambda $}

In order to obtain a better description of the astrophysical
observations at the galactic scale, and also taking into account
the discussion of the previous Section, we may also consider a
first order correction of the metric coefficient $\exp(\lambda) $,
by assuming that $\lambda $ is small. Hence we may approximate the term $\exp(\lambda) $ as $%
\exp(\lambda)\simeq 1+\lambda $, where $\lambda \ll 1$. Moreover,
the condition $5/2 \gg 3\lambda /2-m$ is also satisfied.
Therefore, by neglecting the quadratic term $\lambda^{2}$,
Eq.~(\ref{eqlambda1}) takes the form of an ordinary linear second
order inhomogeneous differential equation, which can be written as
\begin{align}
\left( 1-m\right) ^{2}\frac{d^{2}\lambda }{d\eta ^{2}}-5\left( 1-m\right)
\frac{d\lambda }{d\eta }+2\lambda -4m=0,
\end{align}
and with the general solution given by
\begin{align}
\lambda \left( \eta \right) =2m+C_{+}\exp \left( s_{+}\eta \right)
+C_{-}\exp \left( s_{-}\eta \right) ,
\end{align}
where $C_{\pm}$ are arbitrary constants of integration, and
\begin{align}
s_{\pm}=\frac{5\pm \sqrt{17}}{2\left( 1-m\right) }.
\end{align}

Hence, to first order in both $m$ and $\lambda $, the metric tensor
coefficient $\exp \left( \lambda \right) $ can be represented as
\begin{align}
e^{\lambda }=1+2m+C_{+}r^{s_{+}}+C_{-}r^{s_{-}}.
\end{align}

However, in this case the Riccati equation for $u$, Eq.~(\ref{h1}), cannot
be solved exactly, and numerical methods are needed to further investigate
the physical properties of this solution.

By introducing a new variable $v=d\lambda /d\eta $, then Eq.~(\ref{eqlambda1}%
) can be transformed to a first order differential equation of the form
\begin{multline}
\frac{1}{2}v\frac{dv}{d\lambda }-\frac{1}{1-m} \left(\frac{3}{2}%
e^{\lambda}+1-m\right) v+ \frac{1}{\left( 1-m\right) ^{2}} \\
\times \left( 1-e^{\lambda}\right)^{2} -\frac{1}{\left( 1-m\right) ^{2}}%
\left( 1-e^{\lambda }\right) -2m=0.
\end{multline}

With the help of the transformation $w=1/v$ and by introducing a new
independent variable $\theta =\exp \left( \lambda \right) $, we obtain
\begin{multline}  \label{eqabel}
\frac{dw}{d\theta }+\frac{2}{1-m}\left( \frac{3}{2}+\frac{1-m}{\theta }%
\right) w^{2}-\frac{2}{\theta }\biggl[ \frac{1}{\left( 1-m\right) ^{2}} \\
\times \left( 1-\theta \right)^{2} -\frac{1}{\left( 1-m\right) ^{2}}\left(
1-\theta \right) -2m\biggr] w^{3}=0.
\end{multline}

Hence, Eq.~(\ref{eqlambda1}) has been transformed to a first order nonlinear
second kind Abel differential equation of the form $dw/d\theta =A(\theta
)w^{3}+B(\theta )w^{2}$~\cite{abel}. As it is known from the theory of the
Abel differential equations, an equation of this form has an exact solution
if and only if the condition $d\left[ A(\theta )/B(\theta )\right] /d\theta
=kB(\theta )$ is satisfied, where $k$ is a constant~\cite{abel}. A simple
calculation using the explicit forms of the functions $A(\theta )$ and $%
B(\theta )$ corresponding to Eq.~(\ref{eqabel}) shows that this condition
cannot be satisfied for all values of $m$.

Therefore, it follows that the only exact solution of Eq.~(\ref{eqlambda1})
describing the behavior of the metric coefficient $\exp(\lambda) $ in the
dark matter region in $f(R)$ gravity models to first order in the tangential
velocity is $\lambda =\mathrm{constant}$. To fully investigate the general
behavior of this equation one must use numerical methods.

\section{Discussions and final remarks}

\label{Sec:IV}

From astrophysical observations and from their interpretation in the
framework of the phenomenological Modified Newtonian Dynamics (MOND)
approach~\cite{expl1}, it is known that the acceleration needed to explain
the observed rotation curves is of the order of $10^{-10}\;\mathrm{m}/%
\mathrm{s}^2$, which can be regarded as only a small deviation from general
relativity. Likewise, the Pioneer anomaly, which is also of the order of $%
10^{-10}\;\mathrm{m}/\mathrm{s}^2$, would imply only small modifications of
gravity.

On the other hand, recent laboratory tests have confirmed that
Newton's second law is in ``good agreement'' with accelerations of
the order of $5 \times
10^{-14}\;\mathrm{m}/\mathrm{s}^2$~\cite{PRLa}. Similar
constraints have also been obtained for the inverse square law,
where it was shown that Newton's law holds down to a length scale
of $56\;\mu \mathrm{m}$~\cite {Kapner:2006si}. Hence, it follows
from these observations and experiments that in the $R^n$ modified
theories of gravity the parameter $n$ should be of the form
$1+\epsilon$ with $\epsilon \ll 1$.

In the present paper, we have considered the gravitational field equations
in $f(R)$ modified theories of gravitation in the flat rotation curves
region. Although this is a rather strong assumption, it is valid for at
least a significant region of the total velocity profile. As expected, the
deviations from standard general relativity are only mild and the exponent
of the power law modified gravity takes the form $n=1+v_{\mathrm{tg}}^{2}$.
With the use of Eq.~(\ref{n}) we obtain that this theory is defined by the
action
\begin{align}\label{actionf}
S = \int\! f_0\, R^{1+v_{\mathrm{tg}}^2} \sqrt{-g}\, d^{4}x,
\end{align}
where $f_0$ is a positive constant, given in terms of the
tangential velocity, which can be obtained by using
Eqs.~(\ref{f3}) and~(\ref{f4}), and $v_{\mathrm{tg}}^2$ is a
number of the order of $10^{-6}$. The function $f(R)$ for this
modified gravity model can also be approximated as
$f(R)=f_0R^{1+v_{\mathrm{tg}}^2}\approx
f_{0}R\left(1+v_{\mathrm{tg}}^2\ln R\right)$. Hence the correction
terms to the standard Einstein-Hilbert action is logarithmic.

The weak field limit of the $f(R)$ generalized gravity models has
been discussed recently, for star-like objects, in \cite{Chiba}
and \cite {Fa08}, respectively. By assuming that $f(R)$ is an
analytical function at the constant curvature $R_{0}$, that
$m_{\phi }r\ll 1$, where $m_{\phi }$ is the effective mass of the
scalar degree of freedom of the theory, and that the fluid is
pressureless, the post-Newtonian potentials $\Psi \left( r\right)
$ and $\Phi \left( r\right) $ are obtained for a metric of the
form $ds^{2}=-\left[ 1-2\Psi \left( r\right) \right] dt^{2}+\left[
1+2\Phi(r) \right] dr^{2}+r^{2}\left( d\theta ^{2}+\sin ^{2}\theta
d\varphi ^{2}\right)$. One may then find the behavior of $\Psi
\left( r\right)$ and $\Phi(r)$ outside the star. The analysis
leads to a value of $\gamma =1/2$ for the Post-Newtonian parameter
$\gamma $, which from Solar System observations is known to have a
value of $\gamma =1$. This result rules out most of the $f(R)$
type modified gravity models. However, an analysis of a Lagrangian
of the form given by Eq.~(\ref{actionf}), denoted in \cite{Chiba}
as $f(R)=(R/\alpha )^{1+\delta }$ has also been considered, and
the conclusion is that ``...this analysis is incapable of
determining whether $f(R)=R^{1+\delta }$ gravity with $\delta \neq
1$ conflicts with Solar System tests'' \cite{Chiba}.  Therefore,
the dark matter model obtained from the action given by
Eq.~(\ref{actionf}) is also consistent with the Solar System tests
of general relativity. On the other hand, the results of
\cite{Zakharov:2006uq} also indicate that our model is in
agreement with all observations on scales smaller than $10-20$
kpc.

Another problem facing the $f(R)$ gravity models is the problem of
the stability \cite{Chiba, Fa08,Nojiri:2006ri}. On a time scale of
$\tau \approx 10^{-26}$ s a ``fatal instability'' develops when
$f''(R)<0$ \cite{Fa08}. For the Lagrangian given by Eq.
(\ref{actionf}) we have
$f''(R)=f_0(v_{tg}^2/2)(v_{tg}^2/2+1)R^{v_{tg}^2/2-1}> 0$.
Therefore this type of instability does not develop in the present
model.

The mass discrepancy in clusters of galaxies is a second major
observational evidence leading to the necessity of considering the
existence of dark matter at a galactic and extra-galactic scale.
The total mass of a cluster of galaxies can be estimated in two
ways. First, by taking into account the motions of its member
galaxies, the virial theorem provides an estimate, $M_{V}$.
Second, the total baryonic mass $M$ may be estimated by
considering the total sum of each individual
member's mass. The mass discrepancy arises as one generally verifies that $%
M_{V}$ is considerably greater than $M$, with typical values of
$M_{V}/M\sim 20-30$~\cite{Bi87}.

The virial theorem in $f(R)$ modified gravity was derived, by
using the collisionless Boltzmann equation, in \cite{BoHaLo08}.
The supplementary geometric terms in the modified Einstein
equation provide an effective contribution to the gravitational
energy, and the total virial mass, proportional to the effective
mass associated with the new geometrical term, may account for the
virial theorem mass discrepancy in clusters of galaxies. The
Lagrangian of the modified gravity model, which can be obtained in
terms of quantities directly related to the physical properties of
the clusters, and which can be determined from astrophysical
observations, is again of the form of the action in
Eq.~(\ref{actionf}). This shows that a modified gravity model,
with an action of the form (\ref{actionf}), could give a
consistent geometric description of the properties of the ``dark
matter'' on both galactic and extra-galactic scales. Once the main
physical parameters of matter at the galactic and extra-galactic
scale, like the rotational velocities of the test particles around
galaxies, or the intra-cluster gas temperature or the gas density
profile, are known, the action of the modified gravity model can
be completely obtained from observations, and the
viability/non-viability of the model can be directly tested by
using galactic and cluster of galaxy data, which may also offer an
effective alternative to the Solar System tests.

Another possible physical test of the consistency of our model can
be provided by the study of the deflection of light. The
observational study of the propagation of light at the galactic or
galaxy clusters level and, in particular, the investigation of the
deflection of photons passing through the regions where the
rotation curves of massive test particles are flat, represents one
of the most powerful ways by which one could in principle
constrain $f(R)$ gravity as an alternative dark matter model for
galactic/extra-galactic astrophysical systems. In the flat
velocity curves region, the metric coefficients are given by
Eqs.~(\ref{nu}) and~(\ref{rez1}), respectively. The general
analysis of the lensing in the constant velocity region was
performed in Ref.~\cite{Ein}, and the obtained results can also be
applied to the $f(R)$ models. Therefore, lensing effects can in
principle discriminate between the $f(R)$ gravity and other dark
matter or modified gravity models. By using the results
of~\cite{Ein} it follows that the $f(R)$ gravity model predicts
slightly smaller gravitational lensing effects in the constant
velocity region, as compared to the standard dark matter models.

On the other hand, the resulting deflection angle in the present $f(R)$
gravity model is of the same order of magnitude as the lensing angle in the
standard dark matter model, the isothermal sphere model, which is well
confirmed by observations. This shows that our solution is consistent with
the existing observational data, and that high precision observations may
discriminate between the different classes of models proposed to explain the
motion of test particles around galaxies.

In a series of papers~\cite {Cap2}, using $R^n$ gravity models, a
modified Newtonian potential of the form
\begin{align}  \label{modpotential}
\Phi(r) = -\frac{G m}{2r}
\left[1+\left(\frac{r}{r_c}\right)^{\beta}\right],
\end{align}
was considered, where $m$ is the mass of the particle, $r_c$ a
constant and the coefficient $\beta$ depends on the `slope'
parameter $n$ in the modified action. In the weak field slow
motion approximation $\beta $ can be expressed as
\begin{align}
\beta =
\frac{12n^2-7n-1-\sqrt{36n^4+12n^3-83n^2+50n+1}}{6n^2-4n+2}.
\end{align}

Using the modified Newtonian potential, given by
Eq.~(\ref{modpotential}), it was found that the best fit to $15$
low luminosity rotation curves in $R^n $ gravity is obtained for
$n=3.5$~\cite{Cap2} (somewhat lower values, in particular,
$n=2.2$, were obtained in~\cite{Borowiec:2006qr} and~\cite{Mar1},
respectively). These results seem to suggest that a strong
modification, i.e., a rather large value of $n$ as opposed to
$n=1$, of standard general relativity is required to explain the
observed behavior of the galactic rotation curves. As one can see
from Eq.~(\ref{modpotential}), the potential obtained in this
approach is still asymptotically decreasing, but the corrected
rotation curve, although not flat, is higher than the Newtonian
one, thus offering the possibility of fitting the rotation curves
without dark matter. However, as one can see from
Eq.~(\ref{logpot}), the correction term to the Newtonian potential
in the ``dark matter'' dominated region, where the rotation curves
are strictly flat, must have a logarithmic dependence on the
radial coordinate $r$. This correction term does not appear in
Eq.~(\ref{modpotential}), where a power law modified Newtonian
potential is assumed to describe the observed behavior of the
galactic rotation curves. The difference is also related to the
asymptotic behavior of the metric tensor components in the two
models. These differences in the Newtonian limit in the two models
result in different values of the parameter $n$ in the power-law
modified action of the gravity. On the other hand we have to
mention that in our approach we have completely neglected the
effect of the baryonic matter on the space-time geometry.

In conclusion, we have found that regions with exactly flat galactic
rotations curves do not require any kind of dark matter. The rotation curves
are a consequence of the additional geometrical structure provided by the
modified gravity theories, and a very slight modification of the
Einstein-Hilbert Lagrangian may account for the existence of ``dark
matter''. One can, in principle, rewrite the resulting field equations in
terms of the Einstein tensor and interpret the remaining term as a
geometrical energy-momentum tensor. The presence of these higher order terms
seems to provide us with an elegant geometric interpretation of the dark
matter problem.

\acknowledgments We would like to thank the anonymous referee for
comments and suggestions that helped us to improve the manuscript.
The work of CGB was supported by research grant BO 2530/1-1 of the
German Research Foundation (DFG). The work of TH was supported by
the RGC grant No.~7027/06P of the government of the Hong Kong SAR.
FSNL was funded by Funda\c{c}\~{a}o para a Ci\^{e}ncia e a
Tecnologia (FCT)--Portugal through the grant SFRH/BPD/26269/2006.

\end{document}